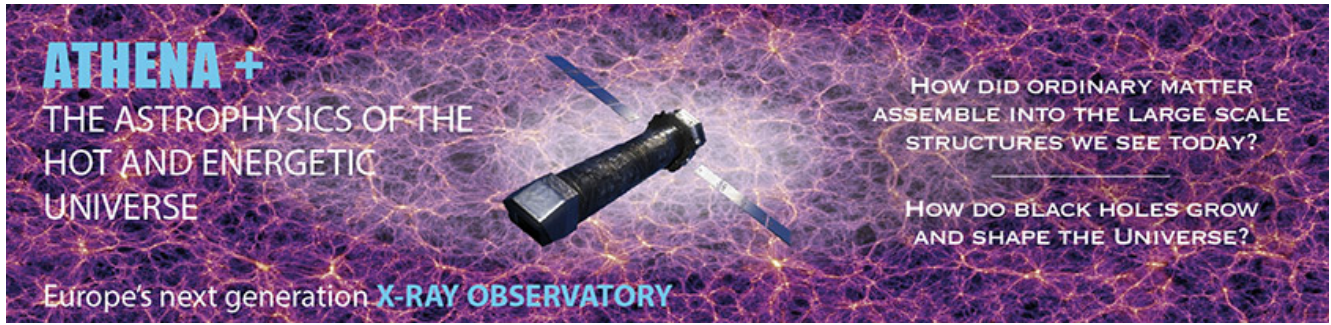

# The Hot and Energetic Universe

An *Athena+* supporting paper

## The evolution of galaxy groups and clusters

## Authors and contributors

**E. Pointecouteau, T. H. Reiprich,** C. Adami, M. Arnaud, V. Biffi, S. Borgani, K. Borm, H. Bourdin, M. Brueggen, E. Bulbul, N. Clerc, J. H. Croston, K. Dolag, S. Ettori, A. Finoguenov, J. Kaastra, L. Lovisari, B. Maughan, P. Mazzotta, F. Pacaud, J. de Plaa, G. W. Pratt, M. Ramos-Ceja, E. Rasia, J. Sanders, Y.-Y. Zhang, S. Allen, H. Boehringer, G. Brunetti, D. Elbaz, R. Fassbender, H. Hoekstra, H. Hildebrandt, G. Lamer, D. Marrone, J. Mohr, S. Molendi, J. Nevalainen, T. Ohashi, N. Ota, M. Pierre, K. Romer, S. Schindler, T. Schrabback, A. Schwope, R. Smith, V. Springel, A. von der Linden.

## 1. EXECUTIVE SUMMARY

By 2028, the cosmological parameters describing the evolution of the Universe as a whole will likely be tightly constrained, e.g., through the eROSITA and Euclid missions. Progress will have been made in understanding the gravitational assembly of structure via the study of the galaxy distribution and evolution (e.g., from Euclid and LSST). However, major astrophysical questions related to the formation and the evolution of galaxy clusters – the largest collapsed structures in which a significant fraction of galaxies is trapped – will still remain:

- What is the interplay of galaxy, supermassive black hole, and intergalactic gas evolution in the most massive objects in the Universe – galaxy groups and clusters?
- What are the processes driving the evolution of chemical enrichment of the hot diffuse gas in large-scale structures?
- How and when did the first galaxy groups in the Universe, massive enough to bind more than $10^7$ K gas, form?

Elements heavier than helium are present at all scales in the Universe, forming the backbone of rocky planets, but are also spread in large quantities at mega-parsec scales in the hot intergalactic medium. The bulk of such elements was likely generated when star formation in galaxies was at its peak (between around redshift 1 and 3), feeding back huge quantities of material and energy into the galactic and intergalactic surrounding media. The evolution of galaxies is tightly coupled to their central super-massive black holes (SMBHs), which also reach the climax of their activity within the same redshift range. Current X-ray observations have revealed that the ensuing feedback from active galactic nuclei (AGN) in central galaxies in galaxy groups and clusters has a major effect on the thermodynamics and heavy element distribution in the hot intra-cluster gas, and impacts the stellar mass of the brightest galaxies in the Universe (those in cluster centres). Thus both feedback and enrichment processes are strongly interleaved and most certainly strongly affect the surrounding inter-galactic medium whilst it is being accreted, heated to tens of millions of degrees, within forming groups and clusters (i.e., $0.5 < z < 2.5$).

The physical and chemical properties of the hot intra-cluster gas in the aforementioned redshift range are a key to understand the impact of galaxy formation and supermassive black hole evolution within their environments. Due to the continuum and line emissions of this medium at X-ray wavelengths, only a large X-ray observatory, such as *Athena+*, combining high sensitivity, very good spectral and spatial resolution, will have the power to lead to major breakthroughs in view of these issues.

Taking advantage of *Athena+'s* throughput and angular resolution, we will constrain the gas density and temperature distribution in groups and clusters out to unprecedentedly high redshifts ($z\sim2$). We will measure of the evolution of structural properties such the entropy and gas fraction distributions. In combination with measurement of the evolution of global scaling relations such as the luminosity–temperature relation, this will allow us to determine how and when energy was deposited into the intra-group/-cluster medium. *Athena+'s* unique spectroscopic capabilities will allow us to measure heavy element abundances via emission lines, and to characterise the enrichment history and mechanisms of the hot plasma in the progenitors of today's massive clusters at $z\sim2$. Finally, a powerful observatory such as *Athena+* will open a large discovery space, which will allow us, for instance, or instance, to potentially discover some of the first collapsed groups harboring X-ray emitting hot gas out to $z\sim3$.

## 2. INTRODUCTION

In the process of hierarchical formation of structure, most galaxies are captured within larger over-densities of matter and assemble as groups. The climax of galaxy group and cluster formation happens between redshifts 0.5 and 2.5, when the star formation activity in galaxies is around its maximum (Hopkins & Beacom 2006, Bouwens et al. 2010) and the central super-massive black holes (SMBH) undergo the peak of their accretion activity (Barger et al. 2004, Merloni et al. 2004). These intense astrophysical processes feed huge quantities of gas, heavy elements, and energy into the increasingly dense and hot intra-group and -cluster gas. At the same time, these massive halos (i.e., with mass larger than a few $10^{13}$ $M_{sun}$) grow through continuous smooth accretion from large filamentary structures (Voit et al. 2003, Voit 2005) and, more violently, through merger events, trapping huge amounts of gas heated to tens of millions of degrees. Therefore, the thermodynamical state of the intra-cluster medium bears the signature of the energy and material fed back by galaxy and SMBH co-evolution, which reciprocally is strongly impacted by the dense and hot surrounding medium.

While our current understanding of the local Universe ($z<0.5$) provides us with a coherent picture of the overall properties of clusters and groups (though physical mechanism ruling their hot baryons still need to be understood in their details), the evolution of halo properties and the physical processes involved in their formation remain unknown to us because of the limitation of current observing facilities. Fundamental questions remain to be answered:

- What is the interplay of galaxy, supermassive black hole, and intergalactic gas evolution in the most massive objects in the Universe – galaxy groups and clusters?
- What are the processes driving the evolution of chemical enrichment of the hot diffuse gas in large-scale structures?
- How and when did the first galaxy groups in the Universe, massive enough to bind more than $10^7$ K gas, form?

Indeed, nearby groups and clusters do not appear as expected in simple gravity-only models (e.g., Cavagnolo et al. 2009, Pratt et al. 2009). Extra energy input, extra physics is required. It is unknown, however, what is the relative role of the two primary candidates responsible for feedback (supernovae and AGN) and how it changes with time. Elements heavier than helium are present at all scales in the Universe, forming the backbone of rocky planets, but are also spread in large quantities at mega-parsec scales in the hot intergalactic medium. Still, we need to understand how exactly the bulk of such elements was created and dispersed in the hot intergalactic medium.

**Answering these questions requires the full physical characterisation of the thermo-dynamical state and chemical composition of the hot intra-group/-cluster medium (ICM) around the time of their formation, i.e., within the redshift range $0.5<z<2.5$.**

With temperatures in the range $10^6$-$10^8$ K, the optically thin intra-group and -cluster gas is invisible at all wavelengths but for the X-ray and mm/sub-mm bands. The X-ray emission is due to continuum thermal bremsstrahlung and atomic line emission of the collisionally ionised gas while the mm/sub-mm emission is due to cosmic microwave background photons scattering off the thermal electrons through the inverse Compton effect (the SZ effect). So, the very same electrons give rise to observable effects in two very different wavelength ranges. However, (i) the emission lines of heavy elements, indicators of the chemical enrichment and its evolution, and also tracers of gas motions impacting group and cluster pressure support, are only accessible to X-rays (i.e., in the [0.1-10] keV band). (ii) Beyond $z>0.5$, groups and clusters extend from a few tens of arcseconds to a few arcminutes; i.e., they are extended but small objects. (iii) Their gas density and temperature structure needs to be determined with high accuracy (i.e., ~10%; the temperature being the limiting factor as its determination requires a spectroscopic analysis) to disentangle the aforementioned effects, from their centre to their outer parts. Only X-ray observations have the abilities to fulfil all three requirements[1]. From X-ray observations, all thermodynamical physical properties about clusters are derived from the measurement of the density (via imaging the surface brightness) and the temperature (from the spectroscopy): entropy, gas mass, total mass, gas fraction, pressure, etc.

The mass function of halos is currently being extensively sampled in the high-mass end out to $z\sim1$ (e.g., with Planck, SPT, ACT, DES, Pan-Starrs). This census will be extended towards less massive systems and out to $z\sim2$, with upcoming survey instruments such as eROSITA, SPT-3G, Euclid, and LSST. By the end of the 2020s, we can assume that a large fraction of all systems with mass greater than about $10^{14}$ $M_{sun}$ out to $z\sim2$ are catalogued and, therefore, a unique pool will be available to design representative, detailed high redshift studies with X-rays that cannot be performed in any other waveband. *Athena+* is optimized to perform this task.

## 3. ENERGY DEPOSITION HISTORY OF THE ICM

Baryonic (i.e., "normal") matter accumulates in potential wells of the dark matter distribution, thereby heated to form the hot gaseous atmospheres of massive halos (e.g., Massey et al. 2007). The gas in groups and clusters of galaxies is heated through shocks and adiabatic compression during the evolution of gravitational potential wells, while galaxy-galaxy and galaxy-ICM interactions modify the morphology of the galaxy content, e.g., through ram pressure stripping. These processes, together with SMBH energy feedback within galaxies, quickly depleted the cluster galaxies of their gas, quenching their star formation whilst they fall deeper within the potential well. Meanwhile, the denser parts of the hot ICM undergo radiative cooling. Some of this cooled gas is believed to be accreted onto the central brightest cluster galaxies (BCGs), ultimately feeding the SMBHs hidden in them, triggering star formation and AGN activity within. In return, the ICM is feedback-heated via the outburst of SMBHs and supernova (SN) driven galactic winds, pushing out heavy element-enriched gas into the ICM and regulating the cooling process.

---

[1] The SZ effect measures the thermal pressure (i.e., the product of density and temperature) integrated along the line-of-sight. Density and temperature cannot be easily disentangled without supporting X-ray observations. While refined spectro-photometry in SZ would clearly help with this, the expected precision of temperature measurements (derived from a 2nd order relativistic treatment of the inverse Compton scattering) will be insufficient to properly address the question.

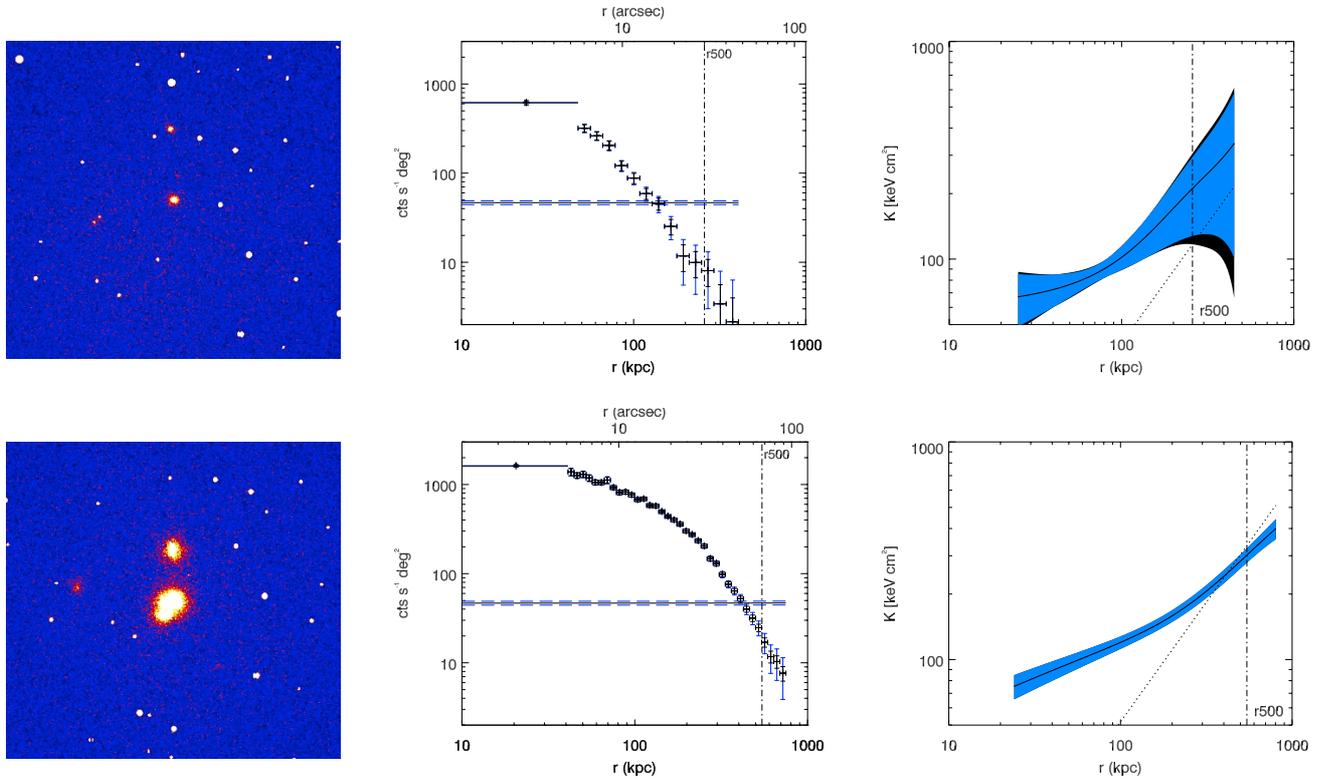

Figure 1: Group of galaxies taken from a numerical simulation, at a redshift of $z$=2 with $M_{500}$=3.5x$10^{13}$ $M_{sun}$ (upper row) and evolved to $z$=1 with $M_{500}$=9.4x$10^{13}$ $M_{sun}$ (lower row). The left column shows 100 ks images folded through the *Athena*+/WFI response, including particle plus X-ray background (including individual AGN sources on the images). The middle column shows the surface brightness profiles, proxies for the gas density profiles, extracted in the [0.5-2.5] keV energy band on the brightest component (after having masked any subcluster). They are recovered out to at least $R_{500}$ in both cases (vertical dot-dashed lines). Statistical errors are shown in black, whilst blue error bars also include 5% systematic errors on the cosmic X-ray background (CXB) flux. The horizontal line shows the total background level (instrumental, CXB and Galaxy). The right column shows the entropy profiles evolving from redshift 2 to 1. They are obtained from an inversion of analytical density and temperature models, given the X-ray surface brightness and projected temperature. Their dispersion reflects the standard deviation of $n_e(r)$ and $k_BT(r)$, and 5% systematic errors on the CXB flux. The dotted lines show the entropy profile derived from numerical simulations that only implement gravitational heating processes (Voit et al. 2005).

**The feedback mechanisms are thus the key to our understanding of galaxy, SMBH and ICM co-evolution in groups and clusters** (e.g., Bouwens et al. 2009, Mullaney et al. 2012, Kravtsov & Borgani 2012). They are likely to provide the extra energy required to keep the cluster cores from cooling all the way down to molecular clouds, to account for the entropy[2] excess observed in the gas of groups and clusters, and to cure the overcooling problem and regulate star formation (e.g., Voit 2005, Croton et al. 2006, Bower et al. 2006, Fabjan et al. 2010, Short et al. 2010, McCarthy et al. 2011). They may also be responsible for the observed correlation between the SMBH mass and the velocity dispersion of their host galaxies (Ferrarese & Merritt 2000; Gultekin et al. 2009), as well as the loss of gas and thus the red sequence shown by elliptical galaxies in dense environments (e.g., Gabor & Davé 2012).

X-ray observations of nearby galaxy clusters provide strong evidence that feedback from SN and/or SMBH affected their overall properties: groups and clusters do not appear as expected in simple gravity-only models, and thus extra energy input and extra physics are required to explain their observed properties. **The history of energy deposited in the ICM is quantified via the entropy distribution within groups and clusters. Entropy is directly derived from X-ray measurements of the surface brightness and temperature of the hot intra-cluster gas.** From the entropy profiles of a wide range of systems spanning a large range of redshift range will allow us to measure: (i) the radial and mass dependence of the entropy excess with respect to pure gravitational collapse; (ii) the central entropy and its evolution (a proxy for cool cores); (iii) the entropy normalisation at $R_{500}$. The key is the interplay of these three quantities over time. How do they evolve with mass and redshift? Do they correlate with other quantities such as star formation rate, AGN activity, etc?

---

[2] "Entropy" here refers to the quantity K=$k_BT$/$n_e^{2/3}$, therefore it is simply defined from the measured electron temperature and electron number density. It is directly related to the standard thermo-dynamic entropy per particle, s = $k_B$ln $K^{3/2}$ + $s_0$, where $s_0$ is a constant (and K∝$e^K$, see Voit 2005).

Current constraints from XMM-*Newton* and Chandra on representative samples of nearby groups and clusters (e.g., Cavagnolo et al. 2009, Sun et al. 2009, Pratt et al. 2010) point towards an excess of entropy (with respect to a purely gravitational model). This excess is mass and radially dependent. It is stronger and extends to larger radii in lower mass systems (and out to beyond $R_{500}$ in the lowest mass systems). These local observational constraints are quite well reproduced by various theoretical models implementing a combination of AGN and wind-driven SN feedback (e.g., Short et al. 2010, Fabjan et al. 2010, McCarthy et al. 2011 – see details in Croston, Sanders et al., 2013, *Athena+* supporting paper). However, their predictions differ strongly at higher redshifts where they cannot be tested with the current generation of X-ray telescopes. It is therefore unknown, which of these two primary candidates (supernovae and AGN) dominates at what epoch. While interaction between central AGN or SN-driven galactic winds and the intra-cluster gas is clearly seen in X-ray images and spectra of clusters (see, e.g., McNamara & Nulsen, 2012), one enigma with invoking SMBH feedback remains: how can a tiny supermassive black hole, the size of our solar system, affect – and even regulate – the largest objects in the Universe: clusters with sizes of $10^7$ light-years?

From gas stirring at the core of groups and clusters to the expulsion of the ICM from their outer parts, we need to understand how the mechanisms of heating and cooling in the ICM shape the features of the halo structures (e.g., the formation of cool cores down to the group regime, quenching of star formation), how they impact the evolution of the gas fraction and how they drive the diffusion of heavy elements at mega-parsec scales? Current X-ray observations neither have the sensitivity to easily probe this evolution beyond redshifts of $z\sim$0.5-0.6 nor out to large radii (e.g., Giodini et al. 2013, Reiprich et al. 2013). XMM-*Newton* and Chandra observations were used to calibrate precisely the local relations (e.g., Arnaud et al. 2005, Vikhlinin et al. 2006, Pratt et al. 2009), however, we must determine the poorly constrained evolution of these relations (e.g., Maughan 2007, Reichert et al. 2011), to achieve the next quantum leap in our understanding. With its large effective area, large field-of-view, and very good spatial resolution, *Athena+* **will perform the required measurements of the global physical properties and their spatial distribution, from local clusters out to $z\sim$2 and down to the group regime** (i.e., objects of characteristic mass $M_{500}\sim5\times10^{13} \, M_{sun}$ and angular size $\theta_{500}\sim30$ arcseconds), where the ICM is more sensitive to the effect of energy input (e.g., Ponman et al. 2003, Sun et al. 2009). These capabilities are illustrated in Fig. 1. *Athena+* will dramatically shift the frontiers of our observations of the ICM in low mass clusters to the redshifts where feedback processes were most active.

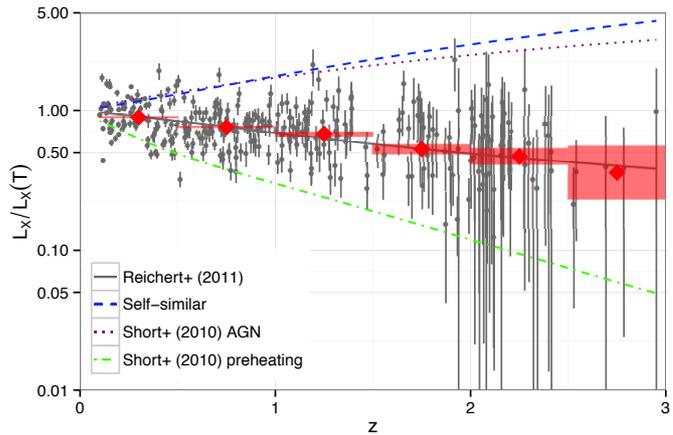

Figure 2: Simulations showing the ratio of the luminosity to the luminosity predicted (by a non-evolving L-T relation and for various feedback scenarios) the temperature as both measured by *Athena+*. This effectively shows its ability to measure the evolution of group and cluster physical properties. The plot includes ~300 groups and clusters ($2 \times 10^{13} < M_{500} < 8 \times 10^{14} \, M_{sun}$) spanning the whole range of redshift accessible to *Athena+* ($z<3$), and it illustrates its potential to easily distinguish between evolution models for groups and clusters at various redshifts. Such constraints will be drawn from the observation of dedicated samples and/or from a global survey strategy. For each object a mass is chosen randomly from the REXCESS sample (see Pratt et al. 2009) and then scaled (in mass and size) to the attributed redshift randomly picked from a uniform distribution between 0.1 and 3. Solid lines picture the predictions for evolution for a self-similar model (dashed blue), for the Reichert et al. (2011) best fit model from which our simulations are drawn (black), AGN feedback (dotted purple) and preheating (dotted-dashed green) models from Short et al. (2010).

For instance, scaling local results (Pratt et al. 2009 and Sun et al. 2009), a 100 ks observation with *Athena+*/WFI of a group with $M_{500} \sim 5 \times 10^{13}$ $M_{sun}$ at $z \sim 1.5$ will allow us to recover the surface brightness and the temperature, thus the entropy and the gas fraction, within $R_{2500}$ to 20% precision for an individual group. With deep observations of $\sim 30$ such groups at $z \sim 1.5$, *Athena+* will provide extremely powerful constraints on the mean ICM structure of high-z groups with which to answer the aforementioned questions. Furthermore, *Athena+* will probe over representative samples of groups and clusters (e.g., built from optical/near-infrared selections based on LSST and Euclid), the evolution of scaling relations across a wide range of redshifts (i.e., from 0.5 to $\sim 2$). Figure 2 illustrates how well competing physical feedback/energy input models, degenerate at low redshift, will be ruled out with *Athena+*. Furthermore, serendipitous and/or dedicated deep surveys with *Athena+*/WFI will probe the evolution of the faint end of the X-ray luminosity function out to $z \sim 3$ (see Sect. 4).

## 4. CHEMICAL EVOLUTION OF HALOS THROUGH COSMIC TIME

Early populations of stars have certainly produced the first heavy elements in the Universe, and their diffusion in the intergalactic medium has probably started from early stars and proto-galaxy winds, or quasar outbursts. Most metals (defined as elements heavier than helium) nonetheless likely originated from the intense period of star formation activity, which occurred over the range of redshift of $z \sim 1$-3 (e.g., Bouwens et al. 2009). Natural outcome of supernova explosions, Si to Ni are mainly produced by type Ia supernovae (SNIa), while the lighter elements from O to Si are produced in core-collapse supernovae (SNcc). N, and to a lesser extent C and F, are produced mainly by Asymptotic Giant Branch (AGB) stars (see Werner et al. 2008 for a review). Metals play a capital role in the thermo-dynamical balance of most astrophysical systems (from planets to stars and galaxies) as they sustain the cooling of their environment by means of emission spectral lines. Metallicity is therefore a key ingredient to the formation and the evolution of structures.

The metals enter the ICM via the gravitational action of ram-pressure stripping of in-falling galaxies, merger-induced gas sloshing, and galaxy-galaxy interactions, and the action of the feedback from super-winds in starburst galaxies and the feedback from AGN, which, e.g., quenches star formation in the assembling BCGs, displacing large amounts of metal-rich gas from their inter-stellar medium into the ICM (e.g., Schindler and Diaferio 2008, Fabjan et al. 2010). The gas is in collisional thermal and ionisation equilibrium (apart from specific regions such as immediately downstream of shocks), and its high temperature prevents the metal atoms to be locked up in dust grains. Therefore, as closed systems, massive halos retain their metals. While the bulk of this activity happened earlier on, the hot intra-group and -cluster gas becomes a fossil record of the peak of metal production by early stellar populations in the history of the Universe.

The global picture described above needs nonetheless to be investigated in order to understand the actual role, scale and impact of each suspected process, and thus the following question needs to be answered: **What are the processes driving the evolution of chemical enrichment of the hot diffuse gas in large-scale structures?** It can, in fact, be broken up into sub-questions: (i) What is the relative contribution of different supernova types (SNIa and SNcc) and stellar winds? How does this evolve with time? (ii) What is the evolution of the heavy element yields of the different supernova source types, if any? (iii) What is the initial stellar mass function (IMF) in (proto-) clusters and does it depend on the environment? (iv) Once produced in stars and supernovae, how are the heavy elements distributed into the hot intergalactic gas in groups and clusters (i.e., at mega-parsec scales)?

These questions can be answered by measurements of the evolution of the abundances of heavy elements, including O, Ne, Mg, Si, S, and Fe, as a function of environment, e.g., centres of galaxy clusters versus outskirts of galaxy groups, and in particular as a function of redshift. In detail, since the various sources (e.g., SNIa, SNcc, AGB stars) synthesize heavy elements in different proportions (yields), their relative role can directly be constrained by abundance ratio measurements (e.g., of O/Fe and Si/Fe). The same measurements will also improve the constraints on the theoretical yields. Similarly, as the IMF enters in the calculation of the integrated yields, the detailed abundance ratio measurements will show whether IMF modifications/evolution are required. Lastly, measurements of the spatial and redshift distribution of elemental abundances and their ratios will disentangle different proposed processes for enriching the intergalactic gas outside galaxies, e.g., early super-winds or late ram pressure stripping   (see also details provided in Ettori, Pratt, et al., 2013, *Athena+* supporting paper).

**The only way to access the abundances of heavy elements in such a hot gas is the measurement of emission line intensity at X-ray wavelengths, by means of high resolution X-ray spectroscopy.** Outside groups and clusters,

abundance measurements of colder gas at UV/optical/IR wavelengths can be performed (see Kaastra, Finoguenov, et al., 2013, *Athena+* supporting paper). However, one loses the closed box advantage of clusters, while the heavy elements are easily blown out of galaxies, making it difficult to assess the integrated amount of synthesized elements. Complementary to X-ray measures, UV/optical/IR/sub-mm observations will provide estimates of the stellar masses and current star formation rates of cluster and group member galaxies as well as in-falling galaxies, establishing the link between the integrated star formation history probed by *Athena+* and today's stellar content.

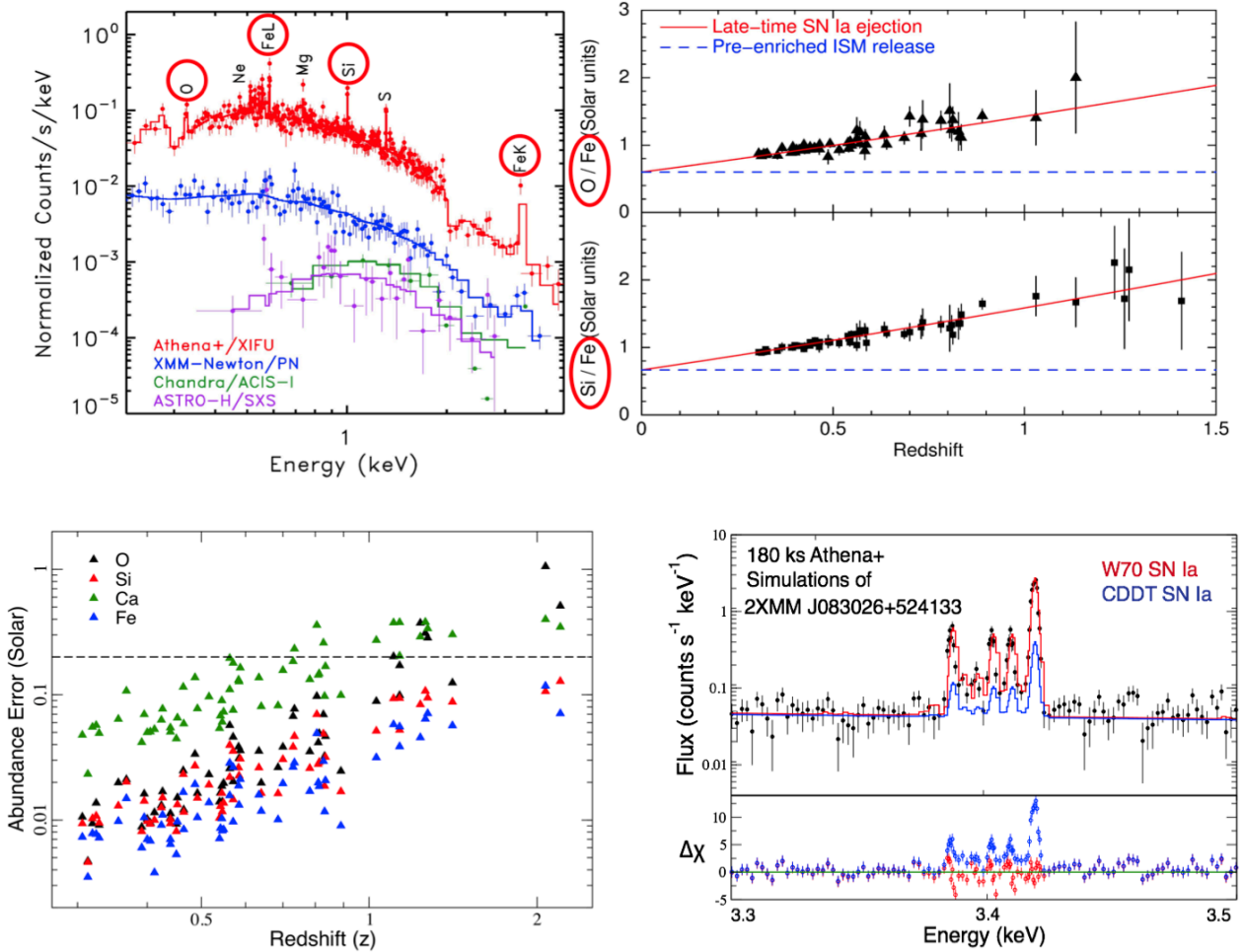

Figure 3: (Top-left) Simulated 150 ks observations (including all background components and instrumental effects) of a *z*=1 cluster with $k_B$T= 3 keV and $L_{X,bol}$=1x10^44 erg/s by *Athena+*/X-IFU, XMM-*Newton*/PN, Chandra/ACIS-I, and ASTRO-H/SXS. Only *Athena+*/X-IFU has the power to spectroscopically resolve the individual lines at this redshift for abundance measurements. (Top-right) O/Fe and Si/Fe ratio for 100 ks *Athena+* exposures of distant clusters in the Balestra et al. (2007) sample as a function of redshift. These abundance ratios will discriminate between a model where recent enrichment is caused by delayed SNIa ejection (red; assumed for these simulations), producing at lower redshift relatively more Fe than O and Si, and a model where the enrichment is due to ISM stripping of already pre-enriched member galaxies (blue), which do not show evolution in redshift. The best-fit model will determine the star formation history of clusters. (Bottom-left) Expected statistical errors in abundance measurements for the same sample, as a function of redshift for 4 elements (O, Si, Ca, Fe) are shown. It shows *Athena+*/X-IFU can measure the Si/Fe ratio even up to *z*=2. (Bottom-right) Another illustration of *Athena+*/X-IFU capabilities to discriminate scenarios of SN enrichment at high redshift using the X-IFU instrument. The spectrum is a simulation of 2XMM J083026+524133, a *z*=1 massive cluster with $k_B$ T = 8.2 keV and $L_{X,bol}$ =2x10^45 erg/s, using the snapec model (Bulbul et al. 2012). It is compared against two scenarios of different yields patterns (see Iwamoto et al. 1999 and Maeda et al. 2010 for the SN Ia W70 and CDDT models respectively).

To date, XMM-*Newton* and Chandra have provided a few hints about cluster Fe abundance evolution from *z*=1 to the present (e.g., Balestra et al. 2007). These constraints suggest that about half of the metals found in the ICM were released into the intergalactic and intra-cluster media prior to *z*~1. To answer the aforementioned sub-questions, we now need to extend the precise measurement of abundances to all abundant elements of astrophysical importance in groups and clusters out to high redshift. **In practice, *Athena+* will provide observations of clusters at redshifts between *z*=0.5-2 to trace their chemical evolution through cosmic time** making use of the tremendous spectral energy resolution of the X-IFU instrument. We will extend the current measurements by measuring Fe (K and L lines) up to $z$=2 and, additionally, O, Ne, Mg, Si, and S up to $z\sim$1.5. Observations of abundance ratios will show at which rate the ICM was enriched with metals (upper-right panel, Figure 3). As another example, the lower-right panel of Figure 3 illustrates how a deep observation (180 ks with *Athena*+/X-IFU) of a single massive and luminous cluster at $z$=1 (i.e., $M_{500}\sim5\times10^{14}M_{sun}$ and $L_X$=2x10^{45} erg/s, based on 2XMM J083026+524133, Lamer et al. 2008) can separate various models of SN nucleosynthesis with different yield patterns, through temperature and abundance measurements, various models of SN nucleosynthesis with different yield patterns.

**None of the currently operating or planned X-ray missions, apart from *Athena*+/X-IFU, will have the capabilities to perform these measurements with the required precision out to high redshifts (see Figure 3).**

## 5. THE COSMIC WEB IN FORMATION AND EVOLUTION

Almost nothing is known about the thermo-dynamical state of massive structures undergoing gravitational collapse at high redshift. Deepening potential wells trap gas, which starts to heat up via both its gravitational in-fall and by non-gravitational feedback from galaxy evolution (SN-driven winds and SMBH outbursts). These assembling halos are the progenitors of today's massive clusters. The relation between their rapid evolution and their forming ICM is yet unstudied, raising the question: **how and when did the first galaxy groups in the Universe, massive enough to bind more than $10^7$ K gas, form?** Detection of extended hot intergalactic gas, is an unmistakable proof for a galaxy group to be fully collapsed into a deep gravitational potential well. To date, XMM-*Newton* and Chandra have detected a few bona fide groups and clusters beyond $z$=1 and up to $z\sim$1.5-2 with fluxes as low as 5x10^{-16} ergs/s/cm^2 (e.g., Bielby et al. 2010, Gobat et al. 2011, Tanaka et al. 2010, 2013, Erfanianfar et al. 2013); though a characterization beyond mere detection requires enormous time investments.

 At even higher redshifts, larger than about $z$=2.5, the detection of the hot gas emission from faint and small groups ($M\sim5\times10^{13}$ $M_{sun}$) is accessible only to powerful X-ray telescopes such as *Athena*+[3]. Indeed, their characteristic radius ($R_{500}$) is about half an arcmin, which is significantly larger than the few arcsecond PSF of the *Athena*+ X-ray telescope (even off-axis, where the PSF's HEW is expected to be <10 arcsec). Extrapolating low-redshift scaling relations, one expects to detect about 600 source photons and 2000 background photons within this area (assuming a 100 ks WFI observation and including the typical off-axis vignetting), resulting in a signal-to-noise ratio larger than 10. This simple calculation has been corroborated by detailed simulations and illustrates that **such groups are detectable as extended sources by *Athena*+ and that the key instrumental parameters are large effective area, good spatial resolution over the full (large) field-of-view, and a low background** (see Figure 4). While extended X-ray emission could be due to non-thermal inverse Compton (IC) emission from high-z radio sources, we will be able to differentiate it from thermal emission via X-ray spectroscopy and by cross-correlation with radio surveys (e.g., LOFAR/E-VLA/SKA) to help identify the IC sources or groups of galaxies harbouring an active radio source. Source detection simulations show that the presence of X-ray AGN in the centres of the high-z groups decreases the probability for detection as extended sources by about 50% but only at large off-axis angles (larger than ~15 arcmin).

By the end of the 2020s, we will hunt these evolved groups via two ways: (i) From the zoology of structures logged in by large surveys like Euclid or LSST, good candidates up to about $z\sim$2 will be selected and targeted by *Athena*+. (ii) A few tens of the first groups at $z$>2.5 and up to $z\sim$3 are expected to be discovered in WFI surveys and serendipitously over the mission life time of ten years. Further identification and redshift estimation will be performed by taking advantage of the Euclid and LSST multi-color surveys. Combined with detailed follow-up observations (e.g., E-ELT, JWST, SKA) of their galactic component, this will provide us with a unique insight on their physical properties less than 3 Gyr after the big bang and thus on the early stages of massive halo formation. **This illustrates the huge discovery space that a powerful telescope such as *Athena*+ will open.**

---

[3] Possible future, planned or conceived, ground- and/or space-based SZ instruments might play a complementary role. It appears quite challenging to detect and characterize small high-z groups with masses ~5x10^{13} $M_{sun}$ with SZ instruments, though.

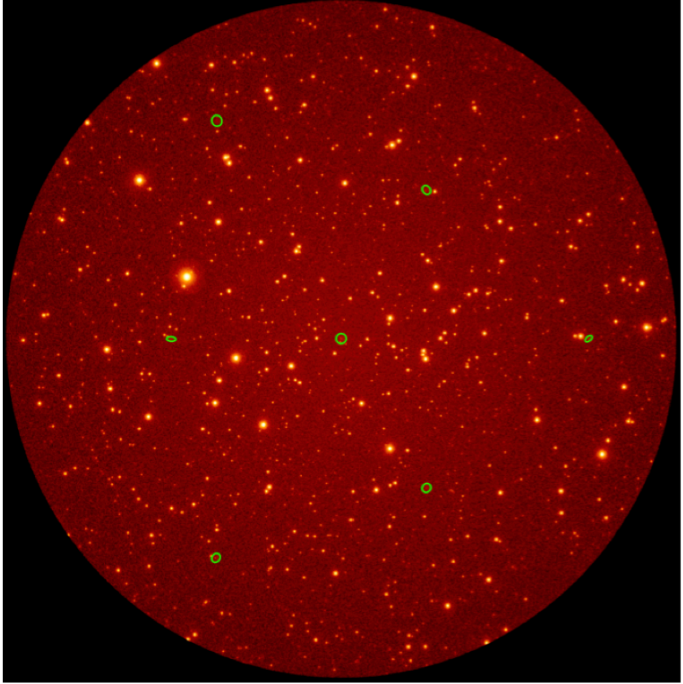

Figure 4: Athena+/WFI 100 ks observation (here shown with 50' diameter). Cosmic X-ray background, Galactic foreground, and particle background are included as well as telescope vignetting and PSF degradation with off-axis angle. The green ellipses show $z$=2.5 groups with $M_{500}$ = 5x10$^{13}$ M$_{sun}$ recovered through a wavelet filtering source detection algorithm (seven such groups are simulated here for illustration). Detailed systematic simulations, taking a variety of physical effects into account, predict that such groups will be detected as extended sources with *Athena+* out to $z\sim3$ (basically no such evolved groups are expected to exist at even higher redshifts).

## 6. CONCLUDING REMARKS

In 2028, the cosmological parameters describing the evolution of the Universe as a whole will be tightly constrained, e.g., through the eROSITA and Euclid missions. However, understanding the physics governing the baryons across cosmic times will still be an open question. A major leap forward in the capabilities of X-ray instruments is needed in order to access the properties of the hot gaseous atmosphere of groups and clusters out to their formation epoch when both star formation and SMBH activity were maximum. This will grant us the ability to reveal how the bulk of heavy elements was generated and distributed in the hot intergalactic medium, and how important AGN and supernova feedback was at high redshift when groups and clusters were born. From extensive detailed observations of low-z systems associated with studies of $z$=1-2 group and cluster properties, a large X-ray observatory such as *Athena+*, combining high sensitivity with excellent spectral and spatial resolution, will bring us a unique view of the evolving hot thermal baryons swarming within collapsing structures. It will bring us the ability to bridge between the era of galaxy formation and early evolution (i.e., beyond $z\sim2$-3; traced by the Ly-alpha forest) and the lower redshift Universe where large-scale structures are assembling and virialising.